# Soliton-similariton switchable ultrafast fiber laser

**Junsong Peng, Li Zhan*, Pan Guo, Zhaochang Gu, Weiwen Zou, Shouyu Luo, and Qishun Shen**

*Department of Physics, Key Laboratory for Laser Plasmas (Ministry of Education), State Key Lab of Advanced Optical Communication Systems and Networks, Shanghai Jiao Tong University, Shanghai, 200240, China*

**Corresponding author: lizhan@sjtu.edu.cn*

For the first time, we demonstrated alternative generation of dispersion-managed (DM) solitons or similaritons in an all-fiber Erbium-doped laser. DM solitons or similaritons can be chosen to emit at the same output port by controlling birefringence in the cavity. The pulse duration of 87-fs for DM solitons and 248-fs for similaritons have been observed. For proof of similaritons, we demonstrate that the spectral width depends exponentially on the pump power, consistent with theoretical studies. Besides, the phase profile measured by a frequency-resolved optical gating (FROG) is quadratic corresponding to linear chirp. In contrast, DM solitons show non-quadratic phase profile. 

Passively mode-locked fiber lasers have attracted much interest owing to wide range of applications. DM soliton lasers have been extensively studied. Besides, similaritons that emerge as a new kind of nonlinear waves have drawn much attention [1]-[2]. They can tolerate strong nonlinearity without wave breaking due to linear chirp. Usually, only one kind of nonlinear wave can be generated in a laser. One question should be addressed is whether different types of nonlinear waves can be generated in the same cavity. This is not only scientifically interesting but also attractive for applications. Recently, Oktem et al[3] reported a new mode-locking regime, in which, the pulse propagates self-similarly in the gain fiber with normal dispersion, and evolves into a soliton in the rest of the cavity.

In this letter, we present a DM soliton-similariton switchable laser, which can output DM solitons or similaritons alternatively at a single output port. Different from the result that they existed in different positions of the cavity,[3] our laser outputs the different pulses at the same port. Once one kind nonlinear wave is generated, the laser can be switched to emit another wave through tuning polarization controllers (PCs). Furthermore, our laser is based on all-fiber devices, which is attractive due to its compact size and easy-operation.

Figure 1 shows the configuration of our ring laser. The cavity is made of a 130 cm EDF (80 dB/m absorption ratio at 1530 nm), which is pumped by a 976 nm laser diode through a wavelength division multiplexer (WDM) pigtailed by 30 cm Nufern 980 fiber, a 270 cm SMF. The GVD parameters of these fibers are -51, 4.5, and 18 ps/(nm·km), respectively. A 10% optical coupler (OC) after the EDF is used to output the signal. A polarization dependent isolator (PDI) sandwiched with two PCs (PC1 and PC2) is used as the mode-locking component.

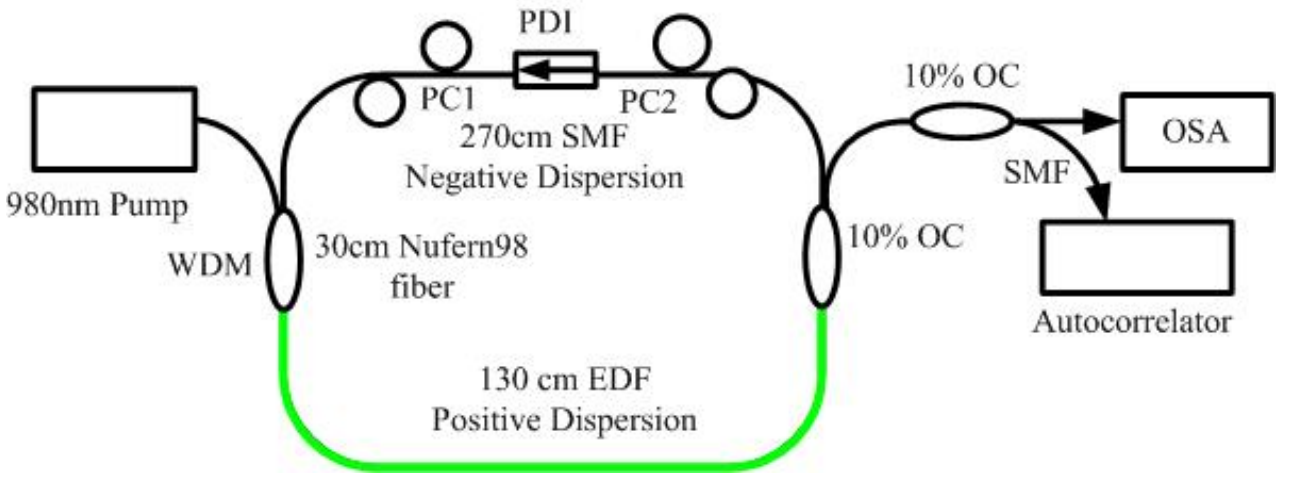

Fig. 1. Experimental configuration of the laser.

The laser cavity is a dispersion-managed one with net dispersion of 0.021 ps$^2$. Both DM solitons and similaritons have been demonstrated around this value of net dispersion[3], [4]. Thus, it's possible to generate these two kinds of waves in a single laser cavity. In terms of net dispersion, it seems no difference between DM solitons and similaritons. What makes them generate respectively in a single laser? It is the filter in the cavity that promises the transition between the two kinds of pulses. In our laser, the filter is invisible, which is formed by fiber birefringence and polarization dependent isolator[5]. The fiber birefringence here is classified as bending-induced birefringence, which is from the fiber wrapped on the PCs[6]. Thus, the bandwidth of the filter can be tuned by rotating PCs[5, 7, 8], as the birefringence magnitude is changed with PCs rotation. Recently, this kind of filter is employed to generate dissipative solitons (DSs) that need a filter to offer additional amplitude modulation[5, 7, 8], which simplifies the construction of DS lasers. The formation mechanism of similaritons in our laser is similar to ref. [3],[9, 10]. Similaritons can be generated by putting an initial pulse in an amplifier with normal dispersion. Remarkably, the output characteristics of similaritons only depend on the energy of incident pulses, and are independent of the initial intensity profiles of the pulses[1]. However, similariton generation becomes a challenge in a laser as the evolution of their characteristics is monotonic, which can't be self-consistent in a laser cavity. Therefore, some mechanism to reverse the evolution must be added. A filter can compensate the evolution, which has been demonstrated.[3, 10] This kind of similariton is called the amplifier similariton. The gain fiber and the filter is the key component to form a nonlinear attractor to generate similaritons, which dictate the pulse evolution. It should be noted that the formation of the amplifier similariton is independent of the net dispersion. In other words, similaritons can be even generated in net anomalous dispersion cavity as long as the normal dispersion gain fiber and a filter is employed, which has been demonstrated very recently[9]. Thus, it can be inferred that dispersion-managed cavities can generate amplifier similaritons if a filter and normal dispersion gain fiber is employed. DM solitons can be observed if the filter is removed or the filter effect becomes weak as only dispersion and nonlinearity is needed to support their generation. Hence, similaritons and DM solitons can be switched depending on the filter bandwidth. In other words, similaritons can be generated when the filter effect is obvious, i.e., the filter bandwidth is small. DM solitons emit when the filter effect is weak, i.e., the bandwidth is wide. It is interesting to consider the counterpart in all normal dispersion cavities. DSs exist in all normal dispersion cavities. Remarkably, the transition between DSs and similaritons was implied in a numerical study, realized by tuning the bandwidth of a filter[11].

Mode-locking was initiated by NPR, and adjusting the PCs was to generate stable pulses. First, DM solitons are observed. Fig. 2(a) shows the optical spectrum of the pulses with 3 dB width of 57 nm. The black curve shows the measured spectrum and the red one is Gaussian fitting. Such a wide spectrum width and spectrum shape are typical ones in DM soliton lasers when the net dispersion is around zero[12-14]. It is well known that the spectrum shape of DM solitons is Gaussian type. The spectrum in Fig. 2(a) is so wide that it exceeds the gain bandwidth of EDF, thus it deviates a little from Gaussian fitting. This deviation is typical in nearly zero net dispersion DM soliton lasers[13, 14]. Due to dispersion variation in the cavity, no Kelly bands were found[4]. The repetition rate of the pulse trains is 47.6 MHz corresponding to the cavity length of 4.3 m. Fig. 2(b) is the autocorrelation trace of DM solitons and Gaussian fitting (red). The duration is 87-fs assuming a Gaussian profile, and the time-bandwidth product is 0.588.

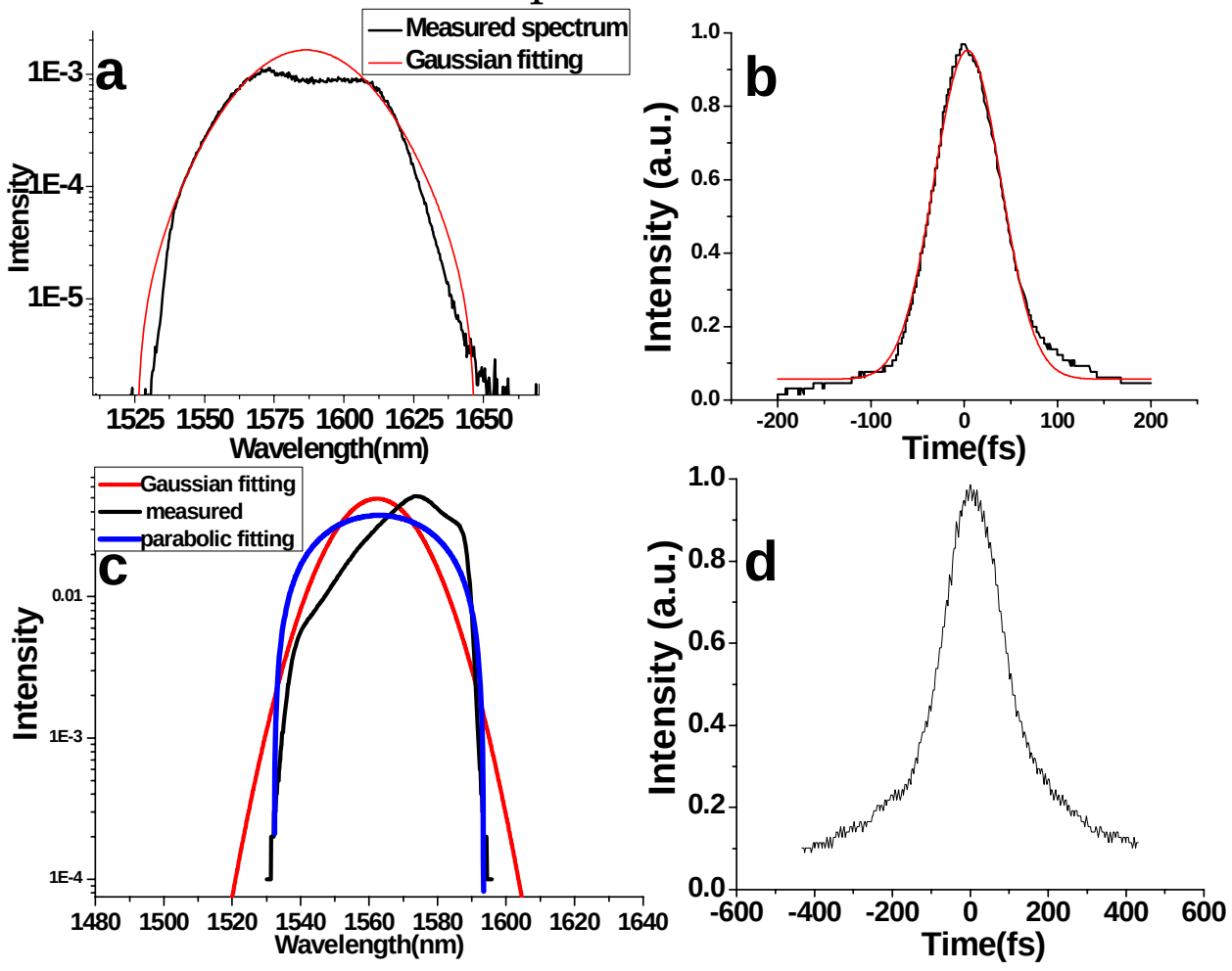

Fig. 2. The spectrum (a) and autocorrelation trace (b) of DM solitons. (c), (d) shows the spectrum and autocorrelation trace of similaritons respectively.

Besides, another mode-locking regime can be observed in this laser. When slightly rotating one PC paddle, the spectrum suddenly becomes narrow, and its two sides are very steep, as shown in Fig. 2 (c). It is well known that the spectrum shape of similaritons is close to parabolic function. The spectrum in Fig. 2 (c) is then fitted by parabolic (blue) and Gaussian (black) functions respectively, and it can be seen that parabolic fitting is closer to the spectrum than Gaussian fitting, especially at pulse edges, indicating that the pulse is a similariton. Fig. 2 (d) shows the autocorrelation trace. The duration is 248-fs assuming a parabolic profile. It should be noted that DM solitons and similaritons emission are very stable. The laser can last for couple of days when it works on either of the two states.

Similaritons can also be distinguished by the fact that the spectral width of similaritons depends exponentially on pump power[15]. In Fig. 3, we experimentally demonstrate this relationship[15]. It shows that the spectral width depends exponentially on the pump power as the red line. Spectral width increases 4.2 nm without changes in the spectral shape when pump power increases. The process is reversible. This relationship was also found in ref.[16] to show similariton properties. However, the spectral width of similaritons increases slightly with the pump power in ref. [16], which is limited by the narrow spectral width of similaritons in their laser.

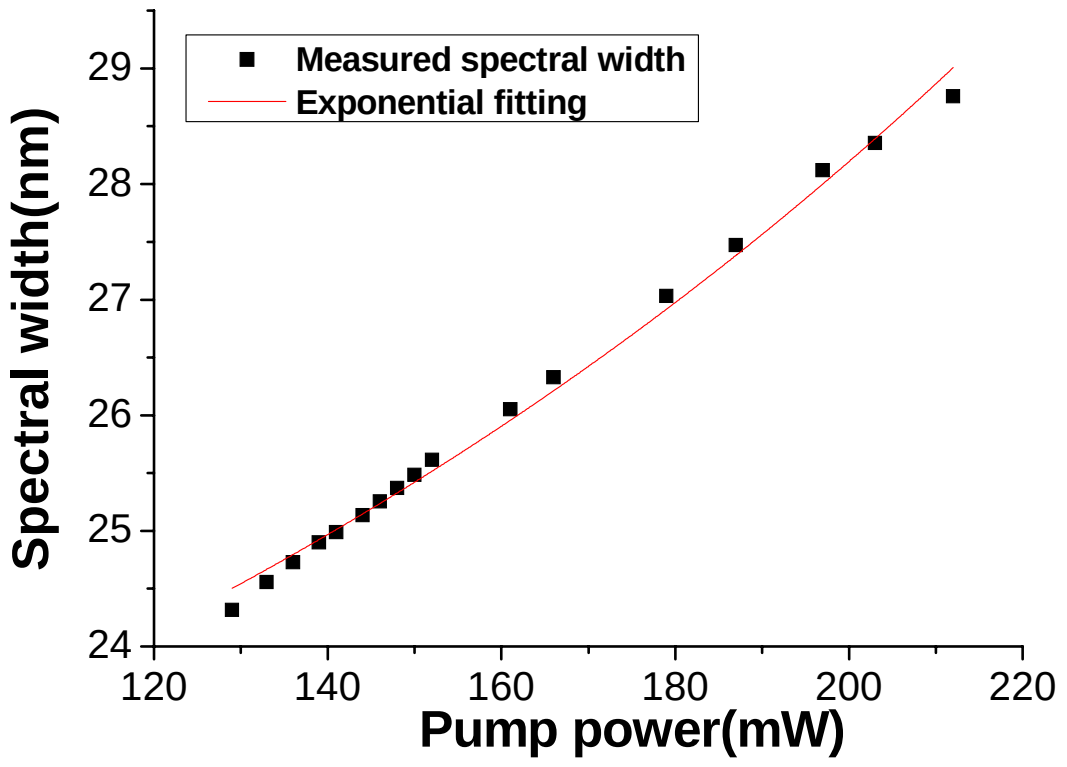

Fig. 3. Spectral widths scaling with pump power.

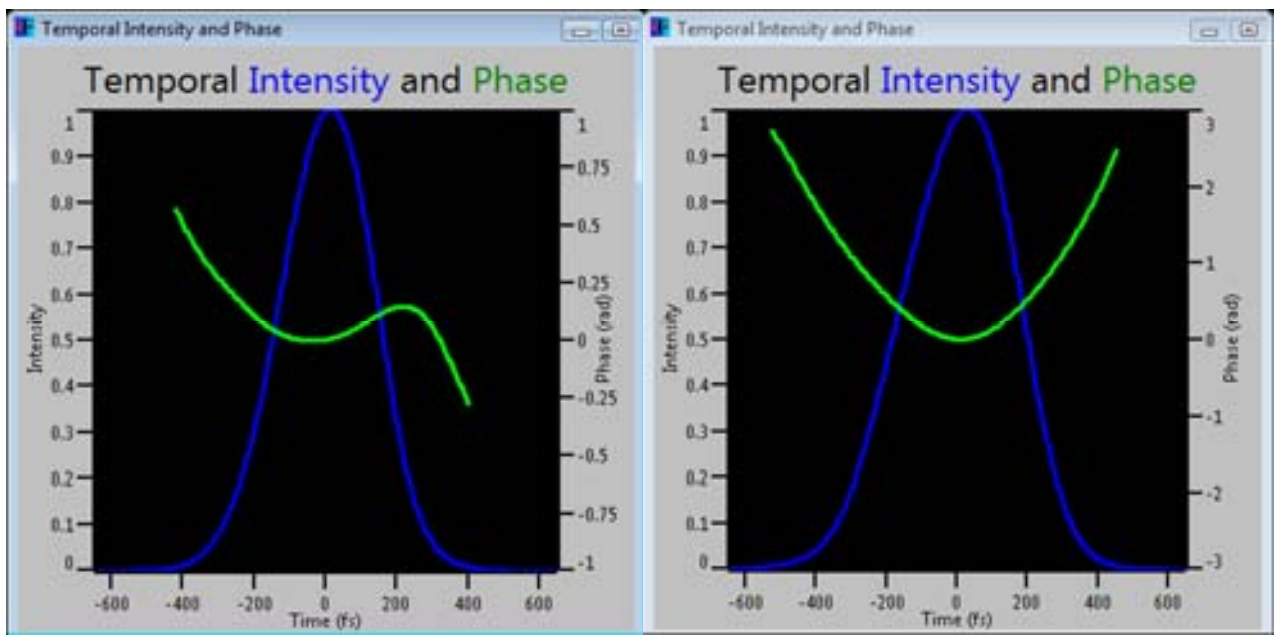

Fig. 4 FROG output of the pulses: DM solitons (left), similaritons (right).

Besides, similaritons can be distinguished from other solitons by a quadratic phase profile, i.e., linear chirp[15, 17]. A FROG is then used to measure the phase of the pulses. Figure 4 (left) shows the phase profile of DM solitons, which is not a quadratic function, i.e., the corresponding chirp is nonlinear. Figure 4 (right) shows that of similaritons. Remarkably, it is very different from that in Fig. 4 (left), which is a quadratic function corresponding to a linear chirp. The pulses are both broaden due to a segment of pigtailed fiber in the FROG. As mentioned, the similariton is generated by a filter in the cavity, hence the filter should affect the spectrum width of similaritons. We examine this relationship by rotating PCs corresponding to changing bandwidth of the filter after similaritons are observed. As shown in Fig. 5, the spectrum width increases (decreases) as one PC paddle rotated clockwisely (counterclockwisely), and the process is reversible, which is consistent with prediction. A similar result is also found in a DS laser[7] in which the spectrum width of DS also varies with tuning PCs.

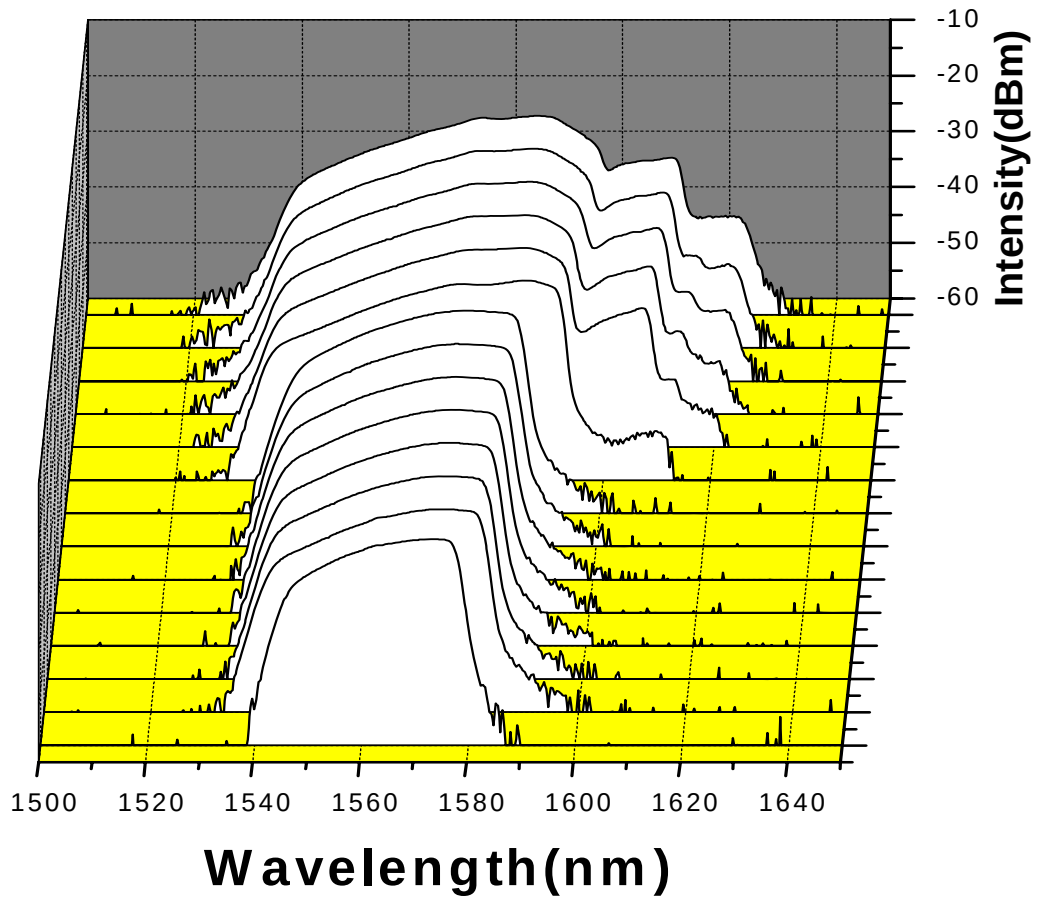

Fig. 5. The spectrum of similaritons scaling with PC rotating.

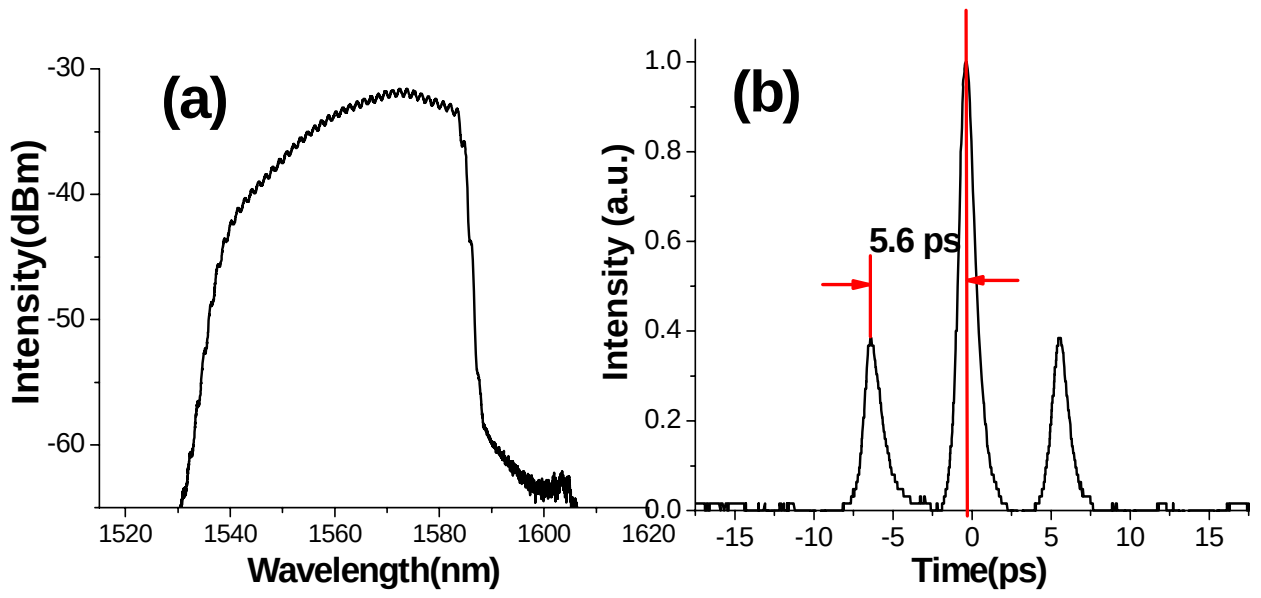

Fig. 6. The spectrum and autocorrelation trace of bound states of similaritons.

Additionally, bound states of similaritons are observed when the pump power increases to 230 mW. Fig. 6 shows the results, in which, the spectrum is modulated and the autocorrelation curve shows three peaks. The period of the modulated spectrum is ~1.4 nm, which corresponds to 5.6 ps separation of two pulses in time domain, as seen from the peaks of the autocorrelation trace in Fig. 6 (b).

Apparently, two kinds of pulses can be chosen to emit in a mode-locked fiber laser. The generation conditions are different for the two pulses. DM solitons existed in Hamilton systems, which are formed by balance between nonlinearity and dispersion. Similariton formation not only depends on dispersion and nonlinearity but also loss effects provided by the filter in the laser, which means similariton is one kind of DSs. Transitions between them indicate that Hamilton and dissipative systems can be switched to each other in a mode-locked laser. Furthermore, the transitions are valuable in practical applications due to different properties of the two pulses. A numerical study is highly desired to describe the transition in details, which is currently under way. Additionally, in all anomalous dispersion regimes, two types of pulses have been found to date. One is Sech soliton, and the other is square soliton[18-20] which is a kind of DSs. Previously, they were generated in different laser systems. Guided by our study, they may be also generated in the same laser.

In conclusion, a DM soliton-similariton switchable all-fiber laser is realized. The pulse duration of 87-fs for DM solitons and 248-fs for similaritons have been observed. The transition between them is because of a filter in the cavity. The bandwidth of the filter depends on fiber birefringence in the cavity. Changing the bandwidth, similaritons or DM solitons can be alternatively generated.

This work was supported from the National Natural Science Foundation of China under Grants 61178014 and 11274231, and the key project of the Ministry of Education of China under Grant 109061.

## References

1. M. E. Fermann, V. I. Kruglov, B. C. Thomsen, J. M. Dudley and J. D. Harvey, Phys. Rev. Lett. **84**, 6010 (2000).
2. A. C. Peacock, R. J. Kruhlak, J. D. Harvey and J. M.Dudley, Opt. Commun. **206**, 171 (2002).
3. B. Oktem, C. Ulgudur and F. O. Ilday, Nat Photon **4**, 307-311 (2010).
4. K. Tamura, L. E. Nelson, H. A. Haus and E. P. Ippen, Appl. Phys. Lett. **64**, 149 (1994).
5. L. M. Zhao, D. Y. Tang, X. A. Wu and H. Zhang, Opt. Lett. **35**, 2756 (2010).
6. R. Ulrich, S. Rashleigh and W. Eickhoff, Opt. Lett.**5**, 273 (1980).
7. J. S. Peng, L. Zhan, Z. Gu, K. Qian, X. Hu, Y. Luo and S. Shen, IEEE Photon. Technol. Lett. **24**, 98 (2012).
8. C. Ouyang, P. Shum, K. Wu, J. H. Wong, X. Wu, H. Lam and S. Aditya, IEEE Photonics Journal, **3**, 881 (2012).
9. W. H. Renninger, A. Chong and F. W. Wise, Opt. Express**19**, 22496 (2011).
10. W. H. Renninger, A. Chong and F. W. Wise, Phys. Rev. A **82** 021805 (2010).
11. A. Chong, W. H. Renninger and F. W. Wise, J. Opt. Soc. Am. B **25**, 140 (2008).
12. K. Tamura, E. P. Ippen, H. A. Haus and L. E. Nelson, Opt. Lett. **18**, 1080 (1993).
13. D. Deng, L. Zhan, Z. Gu, Y. Gu and Y. Xia, Opt. Express **17**, 4284 (2009).
14. K. Tamura, E. Ippen and H. Haus, Appl. Phys. Lett. **67**, 158 (1995).
15. V. I. Kruglov, A. C. Peacock, J. D. Harvey and J. M. Dudley, J. Opt. Soc. Am. B **19**, 461 (2002).
16. O. Prochnow, A. Ruehl, M. Schultz, D. Wandt and D. Kracht, Opt. Express **15**, 6889 (2007).
17. F. O. Ilday, J. R. Buckley, W. G. Clark and F. W. Wise, Phys. Rev. Lett. **92**, 213902 (2004).
18. V. Matsas, T. Newson, D. Richardson, and D. Payne, Electron. Lett. **28**, 1391 (1992).
19. W. Chang, J. Soto-Crespo, A. Ankiewicz, and N. Akhmediev, Phys. Rev. A **79**, 033840 (2009).
20. X. Li, X. Liu, X. Hu, L. Wang, H. Lu, Y. Wang, and W. Zhao, Opt.Lett. **35**, 3249 (2010).